\documentclass[a4paper,12pt]{article}
\pdfoutput=1
\usepackage{graphicx, rotating,amssymb,amsmath}
\usepackage{times}
\usepackage{cite}

\ifx\pdfoutput\undefined
\usepackage[dvips,bookmarks]{hyperref}	
\else
\usepackage{hyperref}	
\fi
\hypersetup{colorlinks,bookmarksopen,bookmarksnumbered,citecolor=verdeCP3,
linkcolor=bluCP3,pdfstartview=FitH,urlcolor=rossoCP3}

\def\hhref#1{\href{http://arxiv.org/abs/#1}{#1}} 

\def\be{\begin{equation}}
\def\ee{\end{equation}}
\def\bea{\begin{eqnarray}}
\def\eea{\end{eqnarray}}
\def\ba{\begin{eqnarray}}
\def\ea{\end{eqnarray}}
\usepackage{geometry}

\newcommand{\ft}{{\cal F}_t}
\renewcommand{\(}{\left(}
\renewcommand{\)}{\right)}

\usepackage{bbold}
\usepackage[table]{xcolor}
\usepackage{floatrow}
\newfloatcommand{capbtabbox}{table}[][\FBwidth]
\usepackage{capt-of}

\usepackage{multicol}
\usepackage{color}
\definecolor{rosso}{cmyk}{0,1,1,0.4}
\definecolor{rossos}{cmyk}{0,1,1,0.55}
\definecolor{rossoc}{cmyk}{0,1,1,0.2}
\definecolor{blu}{cmyk}{1,1,0,0.3}
\definecolor{blus}{cmyk}{1,1,0,0.6}
\definecolor{bluc}{cmyk}{1,1,0,0.1}
\definecolor{verde}{cmyk}{0.92,0,0.59,0.25}
\definecolor{verdec}{cmyk}{0.92,0,0.59,0.15}
\definecolor{verdes}{cmyk}{0.92,0,0.59,0.4}

\font\tenrsfs=rsfs10 at 11pt
\font\sevenrsfs=rsfs7
\font\fiversfs=rsfs5
\newfam\rsfsfam
\textfont\rsfsfam=\tenrsfs
\scriptfont\rsfsfam=\sevenrsfs
\scriptscriptfont\rsfsfam=\fiversfs
\def\mathscr#1{{\fam\rsfsfam\relax#1}}

\def\baselinestretch{0.97}

\def\circa#1{\,\raise.3ex\hbox{$#1$\kern-.75em\lower1ex\hbox{$\sim$}}\,}

\newcommand{\beq}{\begin{equation}}
\newcommand{\eeq}{\end{equation}}

\def\circa#1{\,\raise.3ex\hbox{$#1$\kern-.75em\lower1ex\hbox{$\sim$}}\,}
\makeatletter

\def\art{\@ifnextchar[{\eart}{\oart}}
\def\eart[#1]#2#3#4#5#6{{\rm #2}, {#3 #4} {\rm (#6) #5} [{\hhref{#1}}]}
\def\hepart[#1]#2{{\rm #2, \hhref{#1}}}
\newcommand{\oart}[5]{{\rm #1}, {#2 #3} {\rm (#5) #4}}

\newcounter{alphaequation}[equation]
\def\thealphaequation{\theequation\hbox to
0.6em{\hfil\alph{alphaequation}\hfil}}
\def\eqnsystem#1{
\def\@eqnnum{{\rm (\thealphaequation)}}
\def\@@eqncr{\let\@tempa\relax \ifcase\@eqcnt \def\@tempa{& & &} \or
  \def\@tempa{& &}\or \def\@tempa{&}\fi\@tempa
  \if@eqnsw\@eqnnum\refstepcounter{alphaequation}\fi
\global\@eqnswtrue\global\@eqcnt=0\cr}
\refstepcounter{equation} \let\@currentlabel\theequation \def\@tempb{#1}
\ifx\@tempb\empty\else\label{#1}\fi
\refstepcounter{alphaequation}
\let\@currentlabel\thealphaequation
\global\@eqnswtrue\global\@eqcnt=0 \tabskip\@centering\let\\=\@eqncr
$$\halign to \displaywidth\bgroup \@eqnsel\hskip\@centering
$\displaystyle\tabskip\z@{##}$&\global\@eqcnt\@ne
\hskip2\arraycolsep\hfil${##}$\hfil& \global\@eqcnt\tw@\hskip2\arraycolsep
$\displaystyle\tabskip\z@{##}$\hfil
\tabskip\@centering&\llap{##}\tabskip\z@\cr}
\def\endeqnsystem{\@@eqncr\egroup$$\global\@ignoretrue} \makeatother

\long\def\symbolfootnote[#1]#2{\begingroup\def\thefootnote{\fnsymbol{footnote}}
\footnote[#1]{#2}\endgroup}

\definecolor{rossoCP3}{cmyk}{0,.88,.77,.40}
\definecolor{verdeCP3}{rgb}{0.09765625, 0.57421875, 0.1015625}
\definecolor{bluCP3}{rgb}{0, 0.23, 0.67}

\begin{document}

\begin{titlepage}
 \begin{center}
{\Large {\color{rossoCP3}
\bf Strong Dynamics and Inflation: a review} \rule{0pt}{50pt}
}
 \end{center}
 \par \vskip .5in \noindent
\begin{center}
{Phongpichit {\sc Channuie\!\!\symbolfootnote[1]{channuie@gmail.com}}
}
\end{center}
\begin{center}
  \par \vskip .1in \noindent
\mbox{\it
School of Science, Walailak University,
Nakhon Si Thammarat, 80160 Thailand}
   \par \vskip .5in \noindent
\end{center}
\begin{center}{\large \bf \color{rossoCP3} Abstract}\end{center}
\par \vskip .1in \noindent
\begin{quote}
In this article, we review how strong dynamics can be efficiently employed as a viable alternative to study the mechanism of cosmic inflation. We examine single-field inflation in which the inflaton emerges as a bound state stemming from various strongly interacting field theories. We constrain the number of e-foldings for composite models of inflation in order to obtain a successful inflation. We study a set of cosmological parameters, e.g., the primordial spectral index $n_{s}$ and tensor-to-scalar ratio $r$, and confront the predicted results with the joint Planck data, and with the recent BICEP2 data.
\\
[.33cm]
{
\small {{\bf Keywords}: Strongly interacting field theories, Composite Inflaton, Non-minimal coupling, PLANCK and BICEP2 experiments}}
 \end{quote}
\par \vskip .1in
\vfill

 \end{titlepage}

\def\baselinestretch{1.0}
\tiny
\setlength{\unitlength}{1mm}

\hspace{20mm}
\normalsize

\tableofcontents

\section{Introduction}

The underlying theory of inflation constitutes a cornerstone of the standard model of modern cosmology. By definition, it is the mechanism responsible for an early rapid expansion of our universe which is supposed to take place in the very early time. So far, new scalar fields are traditionally used to describe two prominent physics problems, i.e., the origin of mass of all particle in the standard model and cosmic inflation \cite{Starobinsky:1979ty,Starobinsky:1980te,Mukhanov:1981xt,Guth:1980zm,Linde:1981mu,Albrecht:1982wi}. However, the elementary scalar field in field theories is plagued by the so-called hierarchy problem. Commonly, this means that quantum corrections generate unprotected quadratic divergences which must be fine-tuned away if the models must be true till the Planck energy scale. Similarly the inflaton, the field needed to initiate a period of rapid expansion of our Universe, suffers from the same kind of untamed quantum corrections. 

Therefore, finding its graceful exit is one of the great campaigns. Some of the compelling scenarios to solve/avoid the hierarchy problem are, for instance, Technicolor theory (TC) and Supersymmetry (SUSY). On the one hand, the main idea of TC is to introduce a new strongly coupled gauge theory in which Higgs sector of the SM is replaced by a composite field featuring only fermionic matter. On the other hand, one of the prominent motivations of SUSY is to balance the bosonic degrees of freedom with those of the fermionic ones. Here fermions and bosons have partners which will contribute with opposite signs and make the quantum corrections to the Higgs mass very small.

Recently, the claimed detection of the BICEP2 experiment on the primordial B-mode of cosmic microwave background polarization suggests that cosmic inflation possibly takes place at the energy around the grand unified theory scale given a constraint on the tensor-to-scalar ratio. i.e., $r\simeq 0.20$. Since then, a series of papers on model updates has been reviving by this new results. Theses recent efforts include the Higgs-related inflationary scenarios \cite{Nakayama:2014koa,Cook:2014dga,Hamada:2014iga,Germani:2014hqa,Oda:2014rpa}, several paradigms of chaotic inflation \cite{Harigaya:2014sua,Lee:2014spa}, some interesting analyses related to supersymmetry \cite{Harigaya:2014qza,Czerny:2014qqa}, and other compelling scenarios \cite{Ellis:2014cma,Viaggiu:2014moa,Kehagias:2014wza,Kobayashi:2014jga,Hertzberg:2014aha,Ferrara:2014ima,Gong:2014cqa,Okada:2014lxa,Bamba:2014jia,DiBari:2014oja,Palti:2014kza,Kumar:2014oka,Fujita:2014iaa,Chung:2014woa,Antusch:2014cpa,Bastero-Gil:2014oga,Kawai:2014doa,Hossain:2014coa,Kannike:2014mia,Ho:2014xza,Joergensen:2014rya}. 

Nevertheless, the situation is still controversial since some serious criticisms to the BICEP2 results appeared in the literature, e.g., \cite{Mortonson:2014bja}. Furthermore, the Planck collaboration has very recently released the data concerning the polarized dust emission \cite{Adam:2014bub}, while some attempts making a joint analysis of Planck and BICEP2 data have been publicised (see, for example, \cite{Mortonson:2014bja,Cheng:2014pxa}). However, the recent improvement yields the value of $r$ lower than the one initially claimed by Ref.~\cite{Ade:2014xna}.

In this work, we anticipate to solve the cosmological \lq\lq hierarchy problem\rq\rq\, in the scalar sector of the inflation. In doing so, we have posted the compelling assumption that the inflaton needs not be an elementary degree of freedom called the \lq\lq composite inflaton\rq\rq \cite{Channuie:2011rq,Bezrukov:2011mv,Channuie:2012bv,Evans:2010tf} have posted the compelling assumption that the inflaton needs not be an elementary degree of freedom called the \lq\lq composite inflaton\rq\rq\,and remarkably showed that the energy scale of inflation driven by composite inflaton is around the GUT energy scale. Moreover, there has been shown that the composite models of inflation nicely respect tree-level unitarity for the scattering of the inflaton field all the way to the Planck energy scale \cite{Bezrukov:2011mv,Channuie:2012bv} and some efforts have already implemented to study on their phonemenology \cite{Channuie:2013lla,Channuie:2013xoa,Channuie:2014kda}. 

Here we show how strong dynamics can be efficiently used as a viable alternative to study the mechanism of cosmic inflation. In Sec.~\ref{setup} we derive equations of motion to figure out background evolutions. This will allow us to lay out the setup for a generic model of inflation. In Sec.~\ref{theo} we derive $n_{s},\,r$ and ${\cal A}_{s}$ for composite models. In Sec.~\ref{sec:theopre}, we compute the power spectra for the curvature perturbations by using the usual slow-roll approximations and constrain the model parameters of various composite inflationary models using the observational data from Planck and recent BICEP2 observations. Finally, we conclude in the last section \ref{conclusions}.

\section{Composite Setup and background evolutions}
\label{setup}

In this section, we will start by laying out the setup for a generic models of composite inflation. We aim to derive equations of motion to figure out background evolutions and to obtain inflationary expressions. In so doing, we introduce the action for composite models in the Jordan frame (J) in which the inflaton non-minimally couples to gravity taking the form for scalar-tensor theory of gravity as \cite{Channuie:2013lla}
\begin{eqnarray}
\mathcal{S}_{\rm J}=\int d^{4}x \sqrt{-g}\Big[\frac{M^{2}_{\rm P}}{2} F(\Phi) R - \frac{1}{2}G(\Phi)g^{\mu\nu} \partial_{\mu}\Phi\partial_{\nu}\Phi - V(\Phi)  \Big]. \label{action}
\end{eqnarray}
Here $F(\Phi)$ and $G(\Phi)$ in this action are functions of the field $\Phi$ and can be written as
\begin{equation}
F(\Phi) = 1 + \frac\xi{M^{2}_{\rm P}}\,\Phi^{2/D} \,\, {\rm and} \,\, G(\Phi) = \frac{1}{D^2}G_{0}\Phi^{(2 - 2D)/D}\,,
\label{fg1}
\end{equation}
where the composite field $\Phi$ has mass dimension $D$. In the following calculations, we will set $M^{2}_{\rm P} = 1$. The non-minimal coupling to gravity is controlled by the dimensionless coupling $\xi$. Here we introduce a constant $G_{0}$ and $1/D^2$ for later convenience. However, the action under our consideration can practically written in the standard form of the scalar-tensor theory of gravity. To this end, we just redefine the field and write the potential in the form
\begin{equation}
V(\Phi) = \Phi^{4/D} f(\Phi)\quad\quad{\rm with}\quad\quad\Phi \equiv \varphi^{D} \,,
\label{fg}
\end{equation}
where the field $\varphi$ possesses a unity canonical dimension and $f(\Phi)$ can be in general a function of the field $\Phi$. At first glance, the non-minimal term $\xi\Phi^{2/D}R/M^{2}_{\rm P}$ has purely phenomenological origin. It was examined in \cite{Bezrukov:2008ej,Bezrukov:2008ut,DeSimone:2008ei,Barvinsky:2009fy,GarciaBellido:2008ab,Bezrukov:2009db} that with $\xi$ of the order $10^{4}$ the model can produce the spectrum of primordial fluctuation in good agreement with observations. In other words, we can revoke the unacceptable large amplitude of the primordial power spectrum if one takes $\xi=0$ or smaller than ${\cal O}(10^{4})$. According to the above action, the Friedmann equation and the evolution equations for the background field are respectively given by
\begin{eqnarray}
3F H^{2} + 3\dot{F} H  =  3H^{2}F\left(1 + 2{\cal F}_{t}\right) = \frac{1}{2} G \dot{\Phi}^{2} + V(\Phi)\,, \label{h2}
\end{eqnarray}
\begin{eqnarray}
3F H^{2} + 2\dot{F} H + 2F \dot{H} + \ddot{F}=-\frac{1}{2} G \dot{\Phi}^{2} + V(\Phi)\,, \label{h21}
\end{eqnarray}
\begin{eqnarray}
G\ddot{\Phi} + 3HG\dot{\Phi} + \frac{1}{2} G_{\Phi}\dot{\Phi}^{2} + V_{\Phi}= 3F_{\Phi}\left(\dot{H} + 2H^{2}\right)\,,
\label{kg1}
\end{eqnarray}
where ${\cal F}_{t} = \dot{F}/(2HF)$, $H$ is the Hubble parameter, subscripts \lq\lq$\Phi$\rq\rq\,denote a derivative with respect to $\Phi$, and the dot represents derivative with respect to time, $t$. In order to derive the observables, it is common to apply the standard slow-roll approximations such that 
\begin{eqnarray}
|\ddot{\Phi}/\dot{\Phi}|\ll H\,,\quad\quad|\dot{\Phi}/\Phi|\ll H\quad\quad{\rm and}\quad\quad|G\dot{\Phi}^{2}/2|\ll V(\Phi)\,.
\label{kg}
\end{eqnarray}
It is more convenient to work in the Einstein frame (E) instead of the Jordan one. However, the Einstein and Jordan frame are equivalent and related by a conformal transformation of the metric, which amounts to rescaling all length scales. In our presentation below, we will first derive some inflationary parameters in the Einstein frame and then transform to the Jordan one in order to figure out the relation between two frames.

\section{Inflationary Observables}
\label{theo}

The non-minimal coupling between a scalar field and the Ricci scalar may be diagonalized to the minimally coupled system in which the system can basically transformed to the GR form of the action. This approach is well-known as the Einstein frame and is equivalent to the Jordan frame analysis at the classical level. However, it is often more convenient to perform calculations in Einstein frame. Regarding to the frames, there have been some interesting discussions about the Jordan and Einstein frames, see for example \cite{Kaiser:1994vs,Faraoni:1999hp,Catena:2006bd,Artymowski:2013qua}. By performing a conformal transformation, we take the following replacement:
\be
g_{\mu\nu} \longrightarrow \tilde{g}_{\mu\nu} = F\(\Phi\)g_{\mu\nu}\,.
\ee
With the above rescaling replacement, we obtain the action in Eq.~(\ref{action}) transformed into the new frame -- the Einstein frame -- as
\be
\mathcal{S}_{\rm E}=\int d^{4}x \sqrt{-\tilde{g}}\Big[\frac{M^{2}_{\rm P}}{2} \tilde{R} - \frac{1}{2}\partial_{\mu}\chi\partial^{\mu}\chi - U(\chi)  \Big], 
\label{action-e}
\ee
where $\tilde g$ and $\tilde R$ are basically computed from $\tilde{g}_{\mu\nu}$; \lq\lq tildes\rq\rq\,represent the quantities in the Einstein frame, and
\be
\frac{\partial \Phi}{\partial \chi} = \frac F{\sqrt{G F + 3 F_{\Phi}^2 / 2}}\,
\,\,{\rm and}\,\,
U(\chi) = \left. \frac{V(\Phi)}{F^2(\Phi)} \right|_{\Phi = \Phi(\chi)}\,,
\ee
where the subscript denotes a derivative with respect to $\Phi$. We can reexpress inflationary parameters and all relevant quantities in terms of the field $\chi$ if we solve
\be
\chi \equiv \int \frac{\sqrt{G F + 3 F_{\Phi}^2 / 2}}{F}d\Phi\,.
\ee
Using the expression for the slow-roll parameter in the Einstein frame, $\tilde\epsilon$, such that
\be
\tilde\epsilon = \frac 12 \(\frac 1{U}\frac{\partial U}{\partial \chi}\)^2\,,
\label{epsilon-e}
\ee
we can simply obtain the relation between that of two frames, and we see that
\be
\tilde\epsilon = \frac 12 \(\frac{F^2}{V}\frac{\partial \Phi}{\partial\chi}\frac{\partial }{\partial \Phi}\(\frac{V}{F^2}\)\)^2
= \epsilon + \ft\,,
\label{eps-e-eps-j}
\ee
where $\ft \equiv {\dot F}/2HF$; $\epsilon$ is the slow-roll parameter in the Jordan frame given by $\epsilon \equiv \ft - (V_{\Phi}/V)(F/F_{\Phi})\ft$; and the dot denotes a derivative with respect to time, $t$. It is well known that the power spectrum for the scalar perturbation generated from inflaton field $\chi$ in the Einstein frame is given by
\be
{\cal P}_{\zeta} \simeq  \left.\frac{U}{24\pi^2 \tilde\epsilon}\right |_{k |\tau| = 1}\,,
\label{pr-e}
\ee
where the above expression is evaluated at the conformal time $\tau$ when the perturbation with wave number $k$ exits the horizon and the tensor-to-scalar ratio is
\begin{eqnarray}
r  \simeq16\tilde\epsilon\,.
\label{t2s-e}
\end{eqnarray}
Since the power spectra are frame independent, we can use Eq.(\ref{eps-e-eps-j}) to write the power spectrum in Eq.(\ref{pr-e})
and the tensor-to-scalar ratio in Eq.(\ref{t2s-e})
in terms of the Jordan frame parameters as
\ba
{\cal P}_{\zeta} &\simeq& \left.\frac{V}{24\pi^2 F\Big(\epsilon + {\cal F}_{t}\Big)}\right |_{k |\tau| = 1}\,,
\label{pr-cal}\\
r &\simeq&16\left(\epsilon+{\cal F}_{t}\right)\,.
\label{t2s}
\ea
Here, it is convenient (although tricky) to use the results in the Einstein frame, and then we transform the quantities in the Einstein frame into the Jordan one. It is noticed that one obtains the relation between two frames: ${\tilde \epsilon} \Leftrightarrow \epsilon+{\cal F}_{t}$. Having computed the field $\Phi$ at the end of inflation $\Phi_e$  by using the condition $\epsilon(\Phi_{e})=1$, one can determine the number of e-foldings via
\begin{eqnarray}
{\cal N}(\Phi) = \int_{\Phi}^{\Phi_e}\frac{H}{\dot{{\tilde \Phi}}}d{\tilde\Phi}=\int_{\Phi}^{\Phi_e}\frac 1{{\tilde \Phi}'}d{\tilde\Phi}\,,
\label{efold}
\end{eqnarray}
where the subscript \lq\lq $e$\rq\rq\,denotes the evaluation at the end of inflation
and $\Phi'$ is given by 
\begin{eqnarray}
\Phi'= \frac{1}{\left(1+ \frac{3F_{\Phi}^2}{2 F G}\right)}\Big(2 \frac{F_{\Phi}}{G} - \frac{V_{\Phi}}{V} \frac{F}{G}\Big)\,.
\label{epsi-ss}
\end{eqnarray}
Here, we have used the Friedmann equation and the evolution equations for the background field (Eqs.~(\ref{h2})-(\ref{kg1})) and apply the standard slow-roll approximations (Eq.~(\ref{kg})). 
Determining the value of $\Phi$ and $\Phi'$ when the perturbations exit the horizon allows us to compute the spectral index and the amplitude of the power spectrum in terms of the number of e-foldings. The spectral index for this power spectrum can be computed via
\be
n_s = \frac{d\ln {\cal P}_{\zeta}}{d \ln k} + 1 \simeq 1 - 2\epsilon - 2 {\cal F}_{t}
- \Phi'\frac{d\ln(\epsilon+{\cal F}_{t})}{d\Phi}\,.
\label{ns}
\ee
The amplitude of the curvature perturbation can be directly read from the power spectrum and we find
\begin{eqnarray}
{\cal A}_{s}\equiv\log \left[|\zeta|^2\times 10^{10}\right] \simeq \log \left[\left.\frac{V\times10^{10}}{24\pi^{2} F^2\Big(\epsilon + {\cal F}_{t}\Big)}\right.\right]_{c_{s} k|\tau| = 1}\,.
\label{zeta2}
\end{eqnarray}
It is noticed from Eq.~(\ref{ns}) and (\ref{zeta2}) that the spectral index and the amplitude of the curvature perturbation in the Einstein frame respectively reads
\begin{eqnarray}
n_{s}=1-6\tilde{\epsilon}+2\tilde{\eta}+{\cal O}(\tilde{\epsilon}\tilde{\eta},\tilde{\epsilon}^{2},\tilde{\eta}^{2})\quad{\rm with}\quad\tilde{\eta}\equiv \frac{1}{U}\frac{\partial^{2}U}{\partial \chi^{2}}\,,
\label{zetaEi}
\end{eqnarray}
where $\tilde{\eta}$ is the second slow-roll parameter, and
\begin{eqnarray}
{\cal A}_{s}\equiv\log \left[|\zeta|^2\times 10^{10}\right] \simeq \log \left[\left.\frac{U\times10^{10}}{24\pi^{2} \bar{\epsilon}}\right.\right]_{c_{s} k|\tau| = 1}\,,
\label{zetaEiq}
\end{eqnarray}
where the last bit of Eq.~(\ref{zetaEi}) represents the contributions from the second order of inflationary (slow-roll) parameters. It is noticed that the background fields are time-dependent. In order to trust the effective theory during inflation, we need to examine the composite scale compared with the Hubble scale. In so doing, it is convenient to work in the Einstein frame, and we write the Hubble parameter as \cite{Burgess:2009ea}
\begin{eqnarray}
H\equiv \dot{a}/a\simeq \sqrt{U}/M_{\rm P}, \label{Hub}
\end{eqnarray}
where $U$ is the potential in the Einstein frame. To be more explicit, we have reinserted the Planck constant to the above expression. Here, we also have imposed the slow-roll approximation to the equation (\ref{h2}). In the next section, we will examine single-field inflationary models in which the inflaton is a composite state stemming from various four-dimensional strongly coupled theories.  

\section{Theoretical predictions \& observational constraints}
\label{sec:theopre}
In this section, we compute the power spectra for the curvature perturbations by using the usual slow-roll approximations. We will constrain the model parameters of various composite inflationary models using the observational bound for $n_s$ and $r$ from Planck and recent BICEP2 observations, and use ${\cal A}_{s}$ from Planck data.
\subsection{Composite Inflation from Technicolor}

The underlying gauge theory for the technicolor-inspired inflation is the SU(N) gauge group with $N_{f}=2$ Dirac massless fermions. The two technifermions transform according to the adjoint representation of SU(2) technicolor (TC) gauge group, called ${\rm SU(2)}_{\rm TC}$. Here we engaged the simplest models of technicolor known as the minimal walking technicolor (MWT) theory \cite{Sannino:2004qp,Hong:2004td,Dietrich:2005wk,Dietrich:2005jn} with the standard (slow-roll) inflationary paradigm as a template for composite inflation and name it, in short, the MCI model. In order to examine the symmetry properties of the theory, we arrange them by using the Weyl basis into a column vector, and the field contents in this case are
\begin{equation}
{\cal Q}^{a}=\begin{pmatrix}
U^{a}_{L}\\D^{a}_{L}\\-i\sigma^{2}U^{*a}_{R}\\-i\sigma^{2}D^{*a}_{R}
\end{pmatrix}\,, \label{field}
\end{equation}
where $U_{L}$ and $D_{L}$ are the left-handed techniup and technidown respectively, and $U_{R}$ and $D_{R}$ are the corresponding right-handed particles and the upper index $a=1,2,3$ is the TC index indicating the three dimensional adjoint representation. Since the ${\cal Q}$ is four component, the technifermion fields are said to be in the fundamental representation of SU(4). With the standard breaking to the maximal diagonal subgroup, the SU(4) global symmetry spontaneously breaks to SO(4). Such a breaking is driven by the formation of the following condensate:
 \begin{eqnarray}
\left\langle{\cal Q}^{\alpha}_{i}{\cal Q}^{\beta}_{j}\epsilon_{\alpha\beta}{\cal E}^{ij}\right\rangle = -2\left\langle\bar{U}_{R}U_{L}+\bar{D}_{R}D_{L}\right\rangle\,,
\label{v-tc}
\end{eqnarray}
where $i,\,j\,=1,\,.\,.\,.\,,4$ denote the components of the tetraplet of ${\cal Q}$, and $\alpha,\,\beta$ indicate the ordinary spin. The $4\times4$ matrix ${\cal E}^{ij}$ is defined in terms of the 2-dimensional identical matrix, $\mathbb{1}$, as
\begin{equation}
{\cal E}=\begin{pmatrix}
\mathbb{0} & \mathbb{1} \\ \mathbb{1} & \mathbb{0}
\end{pmatrix}\,, \label{field}
\end{equation}
with, for example, $\epsilon_{\alpha\beta}=-i\sigma^{2}_{\alpha\beta}$ and $\left\langle U^{\alpha}_{L}U^{*\beta}_{R}\epsilon_{\alpha\beta}\right\rangle=-\left\langle\bar{U}_{R}U_{L}\right\rangle$. The connection between the composite scalar fields and the underlying technifermions can be obtained from the transformation properties of SU(4). To this end, we observe that the elements of the matrix ${\cal M}$ transform like technifermion bilinears such that 
 \begin{eqnarray}
{\cal M}_{ij}\sim {\cal Q}^{\alpha}_{i}{\cal Q}^{\beta}_{j}\epsilon_{\alpha\beta}\quad{\rm with}\quad i,\,j=1,\,.\,.\,.\,,4\,.\label{mtc}
\end{eqnarray}
The composite action can be built in terms of the matrix ${\cal M}$ in the Jordan frame as \cite{Channuie:2011rq}
\begin{eqnarray}
{\cal S}_{\rm MCI, J}=\int d^{4}x\sqrt{-g}\left[\frac{M^{2}_{\rm P}}{2}R+\frac{1}{2}\xi{\rm Tr}\left[{\cal M}{\cal M}^{\dagger}\right]R+{\cal L}_{\rm MWT}\right]\,,
\end{eqnarray}
where ${\cal L}_{\rm MWT}$ is the Lagrangian density of the MWT sector, see \cite{Channuie:2011rq} for more details. The details of this sector are not relevant for the present discussion. From the above action, the non-minimally coupled term corresponds at the fundamental level to a four-fermion interaction term coupled to the Ricci scalar in the following way:
\begin{eqnarray}
\frac{1}{2}\xi{\rm Tr}\left[{\cal M}{\cal M}^{\dagger}\right]R = \frac{1}{2}\xi\frac{({\cal Q}{\cal Q})^{\dagger}{\cal Q}{\cal Q}}{\Lambda^{4}_{\rm Ex.}}R\,,
\end{eqnarray}
where $\Lambda_{\rm Ex.}$ is a new high energy scale in which this operator generates. Here the non-minimal coupling is added at the fundamental level showing that the non-minimal coupling is well motivated at the level of the fundamental description. However, an instructive analysis of the generated coupling of a composite scalar field to gravity has been initiated in the Nambu-Jona-Lasinio (NJL) model \cite{Hill:1991jc}. With this regard, the non-minimal coupling apparently seems rather natural. Using the renormalization group equation for the chiral condensate, we find
\begin{eqnarray}
\left\langle{\cal Q}{\cal Q}\right\rangle_{\Lambda_{\rm Ex.}}\sim \left(\frac{\Lambda_{\rm Ex.}}{\Lambda_{\rm MCI}}\right)^{\gamma}\left\langle{\cal Q}{\cal Q}\right\rangle_{\Lambda_{\rm MCI}}\,,
\end{eqnarray}
\begin{figure}[!t]
\includegraphics[width=.6\textwidth]{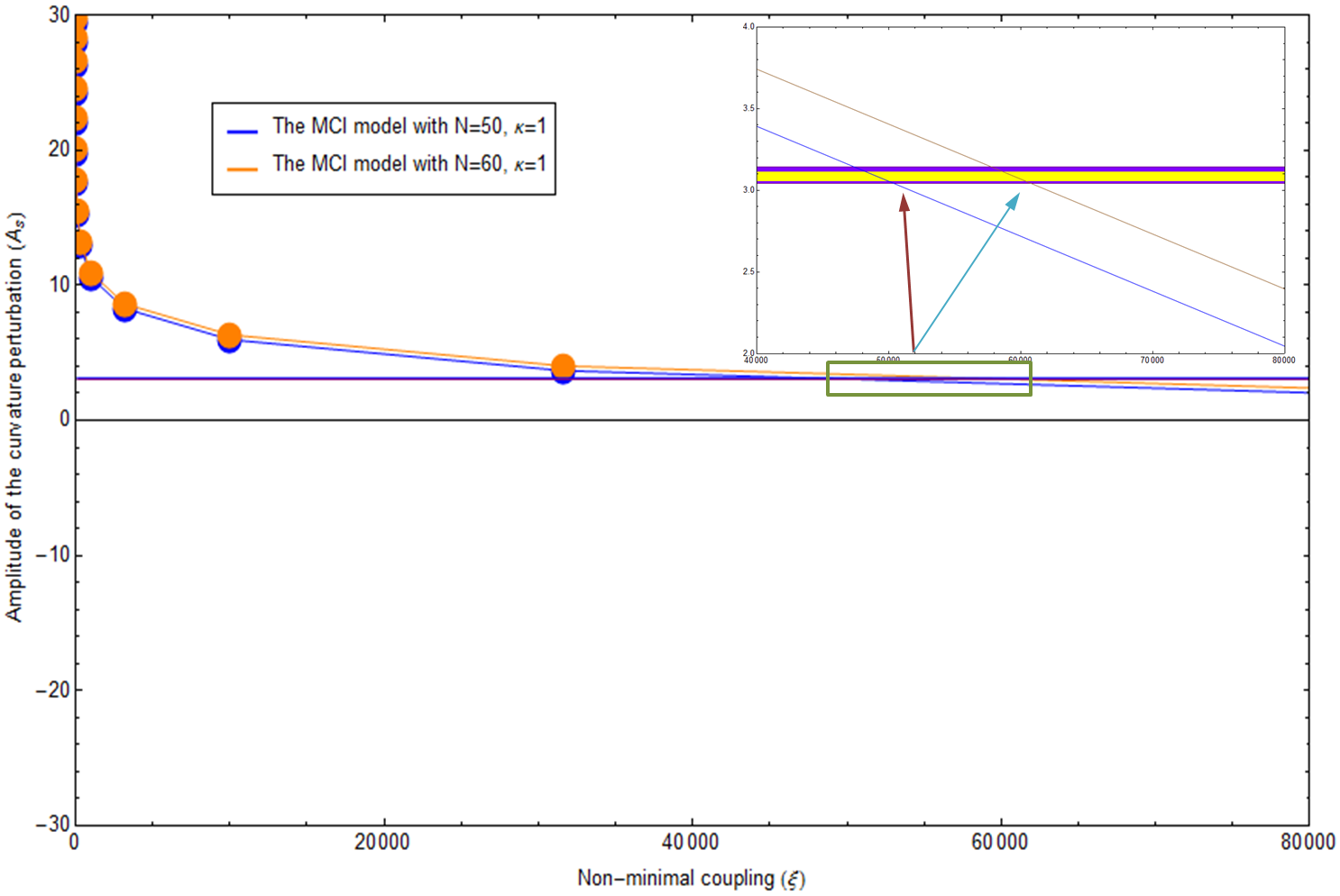}
\caption{The plot shows the relation between the amplitude of the power spectrum ${\cal A}_{s}$ and the non-minimal coupling $\xi$ within a range of $10^{-3}\lesssim \xi\lesssim 10^{6}$ for ${\cal N}=50,60$ predicted by the MCI model. The horizontal bands represent the $1\sigma$ (yellow) and $2\sigma$ (purple) CL for ${\cal A}_{s}$ obtained from Planck.  
}
\label{zetatc}
\end{figure}
where the subscripts indicate the energy scale at which the corresponding operators are evaluated, and basically $\Lambda_{\rm Ex.}\gg\Lambda_{\rm MCI}$. If we assume the fixed value of $\gamma$ is around two the explicit dependence on the higher energy $\Lambda_{\rm Ex.}$ disappears. This is since we have ${\cal M}\sim \left\langle{\cal Q}{\cal Q}\right\rangle_{\Lambda_{\rm MCI}}/\Lambda^{2}_{\rm MCI}$. According to this model at the effective description, the relevant effective theory consisting of a composite inflaton ($\varphi$) and its pseudo scalar partner ($\Theta$), as well as nine pseudo scalar Goldstone bosons ($\Pi^{\mathbb{A}}$) and their scalar partners ($\tilde{\Pi}^{\mathbb{A}}$) can be assembled in the matrix form such that
\begin{eqnarray}
{\cal M}=\left[\frac{\varphi+i\Theta}{2}+\sqrt{2}\left(i\Pi^{\mathbb{A}}+\tilde{\Pi}^{\mathbb{A}}\right)X^{\mathbb{A}}\right]{\cal E}\,,
\end{eqnarray}
where $X^{\mathbb{A}}$'s,\,${\mathbb{A}}=1,...,9$, are the generators of the SU(4) gauge group which do not leave the vacuum expectation value (vev) of ${\cal M}$ invariant, i.e. $\left\langle{\cal M}\right\rangle=v{\cal E}/2,\,v\equiv \left\langle\varphi\right\rangle$. Here the (composite) scale of theory is identified by $\Lambda_{\rm MCI}=4\pi v$, with $v$ the scale of (new) fermion condensate, implying that $\Lambda_{\rm Ex.}\gtrsim 4\pi v$. In this model, the composite inflaton is the lightest state $\varphi$, and the remaining composite fields are massive. This provides a sensible possibility to consider the $\varphi$ dynamics first. In terms of the component fields, the resulting action in the Jordan frame is given by \cite{Channuie:2011rq}:
\begin{eqnarray}
{\cal S}_{\rm MCI}=\int d^{4}x\sqrt{-g}\left[\frac{1 +\xi\varphi^{2}}{2} R - \frac{1}{2}g^{\mu\nu} \partial_{\mu}\varphi\partial_{\nu}\varphi + \frac{m^2}{2}\varphi^2 - \frac{\kappa}{4}\varphi^4  \right]\,,
\end{eqnarray}
\begin{figure}
\begin{center}
\includegraphics[width=.45\textwidth]{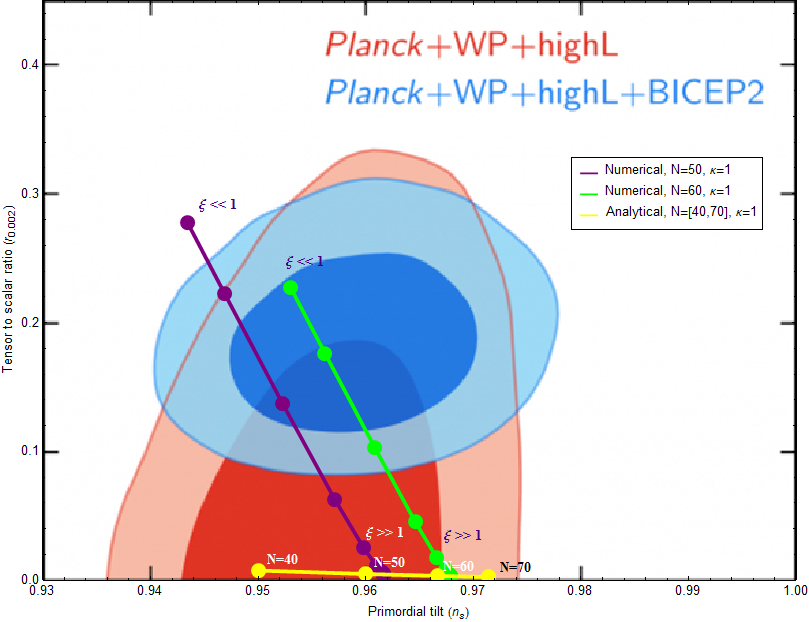}
\includegraphics[width=.488\textwidth]{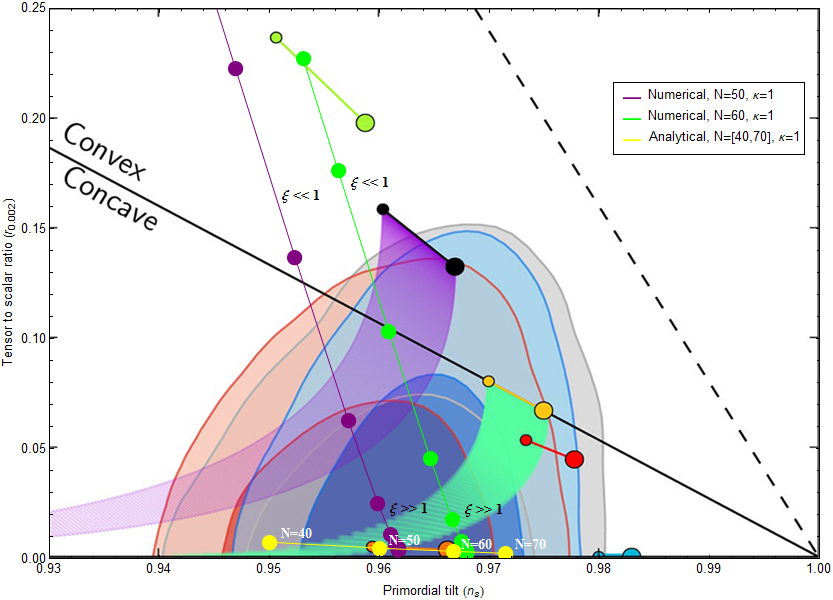}
\end{center}
\caption{The contours show the resulting 68$\%$ and 95$\%$ confidence regions for the tensor-to-scalar ratio $r$ and the scalar spectral index $n_{s}$. Left: The red contours are for the Planck+WP+highL data combination, while the blue ones display the BICEP2 constraints on $r$ \cite{Ade:2014xna}. Right: The figure shows the results from Planck plus various ancillary sets of data \cite{Ade:2013uln}. The plots also show the analytical and numerical predictions given by the MCI model.} 
\label{f1tc}
\end{figure}
in which $\kappa$ is a self coupling and the inflaton mass is $m^{2}_{\rm TI}=2m^2$. In the Einstein frame, the transformed potential reads
\begin{eqnarray}
U(\varphi)=\frac{\kappa}{4}\frac{M^{4}_{\rm P}}{\xi^{2}}\Big(1+\frac{M^{2}_{\rm P}}{\xi\varphi^{2}}\Big)^{-2}\Big(1-\frac{2m^{2}}{\kappa}\varphi^{-2}\Big)\simeq \frac{\kappa}{4}\frac{M^{4}_{\rm P}}{\xi^{2}}\Big(1+\frac{M^{2}_{\rm P}}{\xi\varphi^{2}}\Big)^{-2}\,,
\label{v-tc-e}
\end{eqnarray}
where we have worked in the large field region in which the inflaton is far from its minimum, i.e. $\sqrt{m^{2}/\kappa}$. In the large $\xi$ limit, we obtain $n_s$, $r$ and $|\zeta|^2$ in terms of ${\cal N}$ as
\ba
n_s &\simeq& 1-\frac{8}{3 \varphi^2 \xi} + {\cal O}(1/\xi^{2})
\simeq 1 - \frac 2{{\cal N}}\,,
\label{ns-tc-l}\\
r &\simeq& \frac{64}{3\varphi^4 \xi^2} - \frac{32}{9\varphi^{4} \xi^3} + {\cal O}(1/\xi^{4})
\simeq \frac{12}{{\cal N}^2}\,,
\label{ns-tc-s}\\
|\zeta|^2 &\simeq& \frac{\kappa \varphi^4}{128\pi^2} +\frac{\kappa (-12\varphi^2 + \varphi^{4})}{768\pi^2 \xi}+ {\cal O}(1/\xi^2) 
\simeq \frac{\kappa{\cal N}^{2}}{72\pi^{2}\xi^{2}}\,.
\label{zmci}
\ea
Notice that the above relations lead to the consistency relation, allowing us to write
\begin{eqnarray}
r \simeq  \frac{6}{\cal N}(1-n_{s})\,.
\end{eqnarray}
In this model with $\kappa \sim {\cal O}(1)$, the amplitude ${\cal A}_{s}$ is well consistent with the Planck data up to $2\sigma$ CL for ${\cal N}=60$ and $4.7\times 10^{4}\lesssim \xi \lesssim 5.0\times 10^{4}$, for instance, illustrated in Fig\,(\ref{zetatc}). However, ${\cal A}_{s}$ does strongly depend on ${\cal N}$, and thus the coupling can be lowered (or raised) if ${\cal N}$ changes. We also find for $\xi\gg 1$ that the predictions lie well inside the joint $68\%$ CL for the Planck+WP+highL data for ${\cal N} = [40,\,60]$, whilst for ${\cal N} = 60$ this model lies on the boundary of $1\sigma$ region of the Planck+WP+highL data (the left side of Figure~(\ref{f1tc})). However, with $\xi\gg 1$, the model predictions is in tension with the recent BICEP2 contours (the right side of Figure~(\ref{f1tc})). This is so since the model predictions yield quite small values of $r$. Concretely, the model predicts $\epsilon\sim 1/{\cal N}^{2}$ which no longer holds in light of the BICEPS results for $r=16\epsilon$ such that $r=0.2^{+0.07}_{-0.05}$. Nevertheless, this tension can be relaxed if $\xi$ is very small, i.e. $\xi \sim 10^{-3}$. If this is the case, ${\cal A}_{s}$ cannot satisfy the Planck data unless $\kappa$ gets extremely small, i.e. $\kappa\sim 10^{-13}$. Unfortunately, the prediction with very small $\kappa$ is opposed to the underlying theory. This model predicts $n_{s}\simeq 0.960$ and $r\simeq 0.0048$ for ${\cal N}=50$ with $\xi\gg 1$. Likewise, the Higgs inflation is also in tension with the recent BICEP2 data.

We will complete our discussion in this section by naively clarifying the scales of the theory. It was mentioned in \cite{Channuie:2011rq} that the effective theory for composite inflation cannot be utilized for arbitrary large value of the scalar field, but it rather has some cut-off scale above which the theory is no longer valid. In other words, the theory may in principle produce cross sections that violate unitarity. For MCI, the breakdown of the effective Lagragian happens at $\Lambda_{\rm MCI}=4\pi v$.  To make sure that the effective theory is valid, we impose the condition for which $\varphi<4\pi v$. 

Having imposed the initial value of the inflaton field, $\varphi_{\rm ini}\sim 9M_{\rm P}/\sqrt{\xi}$, it implies that $v\sim (0.81-4.07)\times 10^{16}$\,GeV. This scale is close to the typical grand unification (GUT) scale. The lower bound on the scale of composite inflation arises from having assumed the effective theory to be valid during the inflationary period. The authors of \cite{Channuie:2011rq} have also determined the value of the inflaton field at the end of inflation and found that $\varphi_{\rm end}\sim M_{\rm P}/\sqrt{\xi}$. The constraint on $v$ forbids the identification of the composite inflaton with the composite Higgs. From Eq.~(\ref{Hub}), we can determine the Hubble parameter during inflation and roughly find that $H\lesssim M_{\rm P}/\xi\sim {\cal O}(10^{14})$\,GeV. Apparently, the Hubble scale during inflation is less than all scales we have in this model ensuring the applicability of the effective theory during inflation.

\subsection{Composite Inflation from pure Yang-Mills Theory}

The underlying gauge theory for glueball inflation is the pure SU(N) Yang-Mills gauge theory. The inflaton in this case is the interpolating field describing the lightest glueball. In the same manner with the preceding section, the connection between the composite field and the underlying fundamental description can be also obtained. In this case, the inflaton field is
\begin{eqnarray}
 \Phi=\frac{\beta}{\mathbb{g}}{\rm Tr}\left[{\cal G}^{\mu\nu}{\cal G}_{\mu\nu}\right]\,,
 \end{eqnarray}
where ${\cal G}_{\mu\nu}$ is the standard non-Abelian field strength, $\beta$ is the full beta function of the theory in any renormalisation scheme, and $\mathbb{g}$ is the gauge coupling. We can also demonstrate that the fundamental degrees of freedom are naturally non-minimally coupled to gravity, and features the description at the fundamental level. In doing so, we introduce the non-minimal coupling term as follows:
\begin{eqnarray}
\xi\left(\frac{\beta}{\mathbb{g}}{\rm Tr}\left[{\cal G}^{\mu\nu}{\cal G}_{\mu\nu}\right]\right)^{1/2}R\equiv \xi \Phi^{1/2}R\,,
 \end{eqnarray}
 \begin{figure}
\begin{center}
\includegraphics[width=0.6\textwidth]{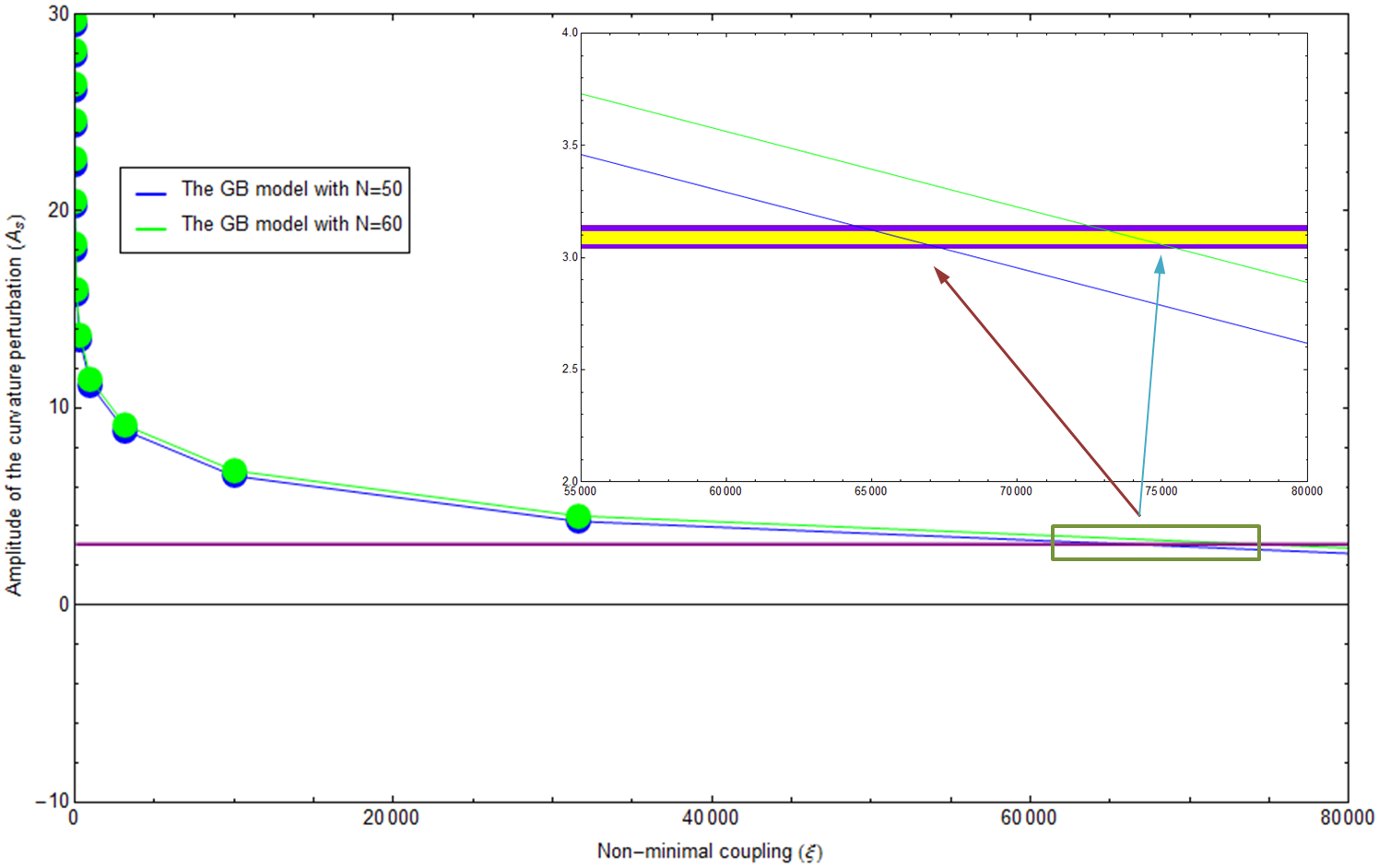} 
\end{center}
\caption{The plot shows the relation between the amplitude of the power spectrum ${\cal A}_{s}$ and the nonminimal coupling $\xi$ with $10^{-3}\lesssim \xi\lesssim 10^{6}$ for ${\cal N}=50,60$ predicted by the GB model. The horizontal bands represent the $1\sigma$ (yellow) and $2\sigma$ (purple) C.L. for ${\cal A}_{s}$ obtained from Planck.} 
\label{f1gb}
\end{figure}
where the $\xi$ is a dimensionless quantity. Here, together with the preceding section, we have explicitly explained how the introduction of the non-minimal coupling is motivated in a natural way with the underlying fundamental descriptions. In this case, the inflaton emerges as the interpolating field describing the lightest glueball associated to a pure Yang--Mills theory. It is worthy to note here that the theory we are using describes the ground state of pure Yang--Mills theory, and of course is not the simple $\phi^4$ theory. For this model, we have
\begin{eqnarray}
 f(\varphi)=2\ln(\varphi/\Lambda)\,,
 \end{eqnarray}
where the glueball condensate scale is parametrised by $\Lambda$. So the effective Lagrangian for the lightest glueball state, constrained by the Yang--Mills trace anomaly, nonminimally coupled to gravity in the Jordan frame reads \cite{Bezrukov:2011mv}
\begin{eqnarray}
{\cal S}_{\rm GB} = \int d^{4}x\sqrt{-g}\Big[\frac{F(\varphi)}{2} R - 16g^{\mu\nu} \partial_{\mu}\varphi\partial_{\nu}\varphi -  2\varphi^{4}\ln\left(\varphi/\Lambda\right) \Big]\,.
\end{eqnarray}
Here we call it, in brief, the GB model. Also the modified version of this model has been considered in \cite{Svendse:2012dva}. In this work, we consider only the large $\xi$ limit and find for this case
\ba
{\cal N} \simeq 3\Big(\ln^2\(\varphi/\Lambda\) - \ln^2\(\varphi_{e}/\Lambda\)\Big) + {\cal O}(1/\xi)\,.
\ea
Here, we can write $\varphi$ in terms of ${\cal N}$ and use Eqs.(\ref{ns}), (\ref{t2s}), and (\ref{zeta2}) to write $n_s$, $r$, and $|\zeta|^2$ in terms of ${\cal N}$. Finally, we obtain for a large $\xi$ limit \cite{Channuie:2013lla}
\begin{eqnarray}
n_s &\simeq&  1-\frac{3}{2{\cal N}} + {\cal O}(\xi)\,,
\label{ns-gb-l}\\
r &\simeq&  \frac{4}{{\cal N}} + {\cal O}(\xi)\,,
\label{t2s-gb-l}\\
|\zeta|^2 &\simeq&  \frac{{\cal N}^{3/2}}{3\sqrt{3}\pi^2\xi^2} + {\cal O}(1/\xi^3)\,.
\label{zeta2-gb-l}
\end{eqnarray}
Notice that the above relations lead to the consistency relation, allowing us to write
\begin{eqnarray}
r \simeq  \frac{8}{3}(1-n_{s})\,.
\end{eqnarray}
We discover that ${\cal A}_{s}$ is well consistent with the Planck data up to $2\sigma$ CL for ${\cal N}=60$ with $7.3\times 10^{4}\lesssim \xi \lesssim 7.5\times 10^{4}$, illustrated in Fig\,(\ref{f1gb}). However, ${\cal A}_{s}$ does strongly depend on the number of e-foldings implying that the coupling can be lowered (or raised) with changing ${\cal N}$. This model provides $n_{s}\simeq 0.967$ and $r\simeq 0.089$ for ${\cal N}=45$ with $\xi\gg 1$.

\begin{figure}
\begin{center}
\includegraphics[width=.45\textwidth]{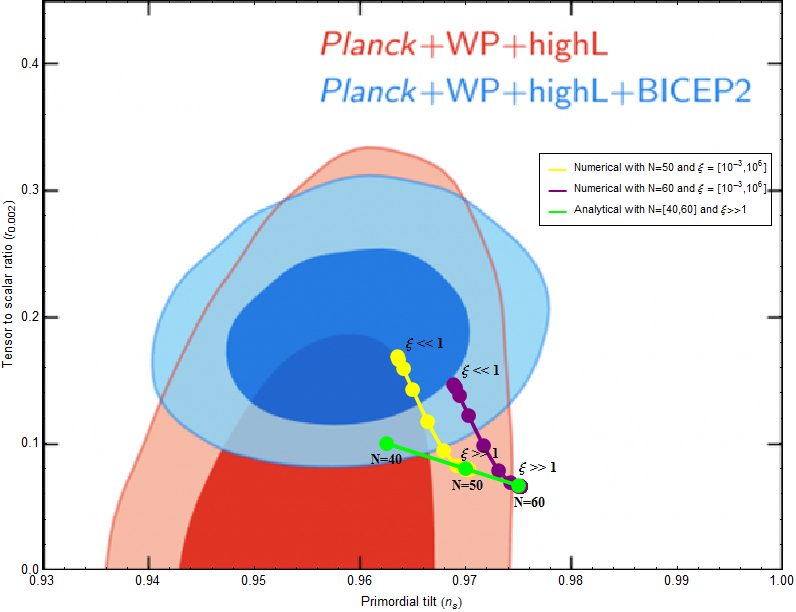}
\includegraphics[width=.488\textwidth]{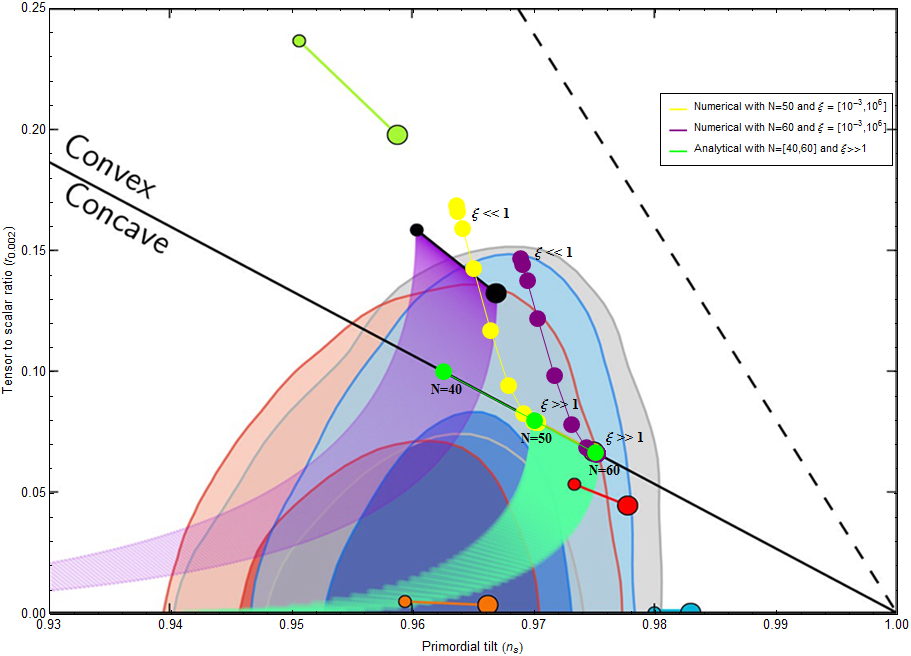}
\end{center}
\caption{The contours show the resulting 68$\%$ and 95$\%$ confidence regions for the tensor-to-scalar ratio $r$ and the scalar spectral index $n_{s}$. Left: The red contours are for the Planck+WP+highL data combination, while the blue ones display the BICEP2 constraints on $r$ \cite{Ade:2014xna}. Right: The figure shows the results from Planck plus various ancillary sets of data \cite{Ade:2013uln}. The plots also show the analytical and numerical predictions given by the GB model.} 
\label{f1togb}
\end{figure}

From the above estimations, we see that when $\xi \gg 1$, $n_s$ , $r$, and $|\zeta|^2$ can satisfy the 95$\%$ C.L. observational bound from Planck data for $50 < {\cal N} < 60$ and $\xi \sim 10^{4}$; see Fig.~(\ref{f1gb}) and \,(\ref{f1togb}). Nevertheless, for such range of ${\cal N}$, $r$ lies outside the $2\sigma$ C.L. with BICEP2 results shown in Fig.~(\ref{f1togb}). The value of $r$ will increase and then satisfy the bound from BICEP2 results when ${\cal N}\lesssim 45$. However, it is obvious that ${\cal N}$ is a model-dependent quantity. However, it is quite subtle if we have ${\cal N}\lesssim 45$ for models of inflation to be viable. This is so since, in order to solve the horizon problem, in the common formulation one frequently uses at least ${\cal N}\subset [50, 60]$. We anticipate this can be further verified by studying the reheating effect. The compatibility between our analytical and numerical results of this model is illustrated in Fig.~(\ref{f1togb}). 

We will complete our discussion in this section by naively clarifying the scales of the theory. As we have seen for the MCI, the effective description of this model would also break down at some energy scale. It was found in \cite{Bezrukov:2011mv} that it is associated with the typical scale for grand unification, $\Lambda > M_{\rm P}/\sqrt{\xi}\sim {\rm a\,few}\,\times\,{\cal O}(10^{16})$\,GeV, in complete agreement with the first model we have earlier mentioned. Here we can trust the effective description up to scales of this order. 

First principle lattice simulations have shown that the fundamental Lagrangian for the pure SU(N) Yang-Mills gauge theory confines at the scale identifiable with $\Lambda$ of the glueball theory. With the result given in \cite{Channuie:2013lla}, i.e. $\varphi_{\rm ini}\sim 88\Lambda$, it was implied that the fundamental description can be used in the perturbative regime to describe the dynamics of the theory at energy scales of the order of $100\Lambda$ and above. For energies below this scale and to describe the vacuum properties of the theory, the effective potential utilised in this presentation works.

Since the standard model couplings are weak at the unification point, whilst the inflationary model is still strongly coupled at this scale (now identified with $\Lambda$). this feature allows us to decouple the contributions of the SM from inflationary theory and ensures that the action formulating inflation does not include any contributions from the SM. For this model, we show that inflation starts at energy scales just below or near the energy scales above which the underlying gauge dynamic is perturbative and expect the perturbative dynamic of the gauge theory to set in before arriving at the Planck scale.  

From Eq.~(\ref{Hub}), we can determine the Hubble parameter during inflation and roughly find that $H\lesssim M_{\rm P}/\xi\sim {\cal O}(10^{14})$\,GeV. Apparently, the Hubble scale during inflation is less than all scales we have in this model ensuring the validity of the effective theory during inflation.
\subsection{Composite Inflation from super Yang-Mills Theory}
\label{sGB}

\begin{figure}
\begin{center}
\includegraphics[width=.45\textwidth]{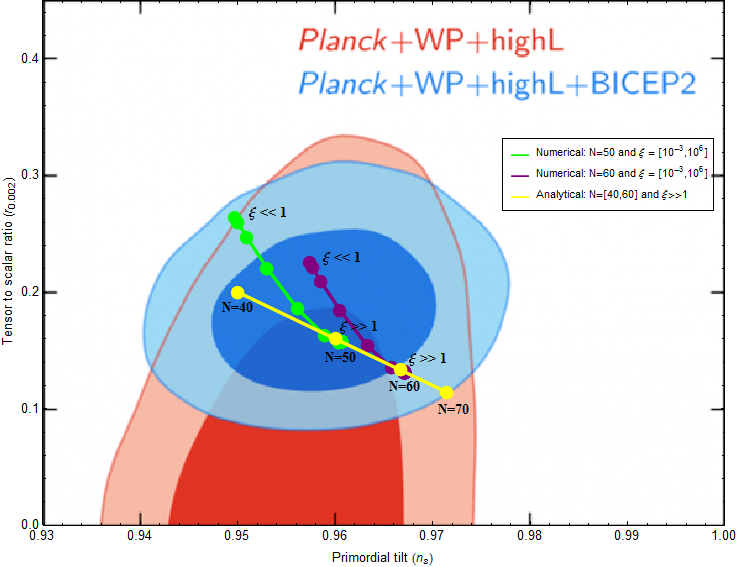}
\includegraphics[width=.488\textwidth]{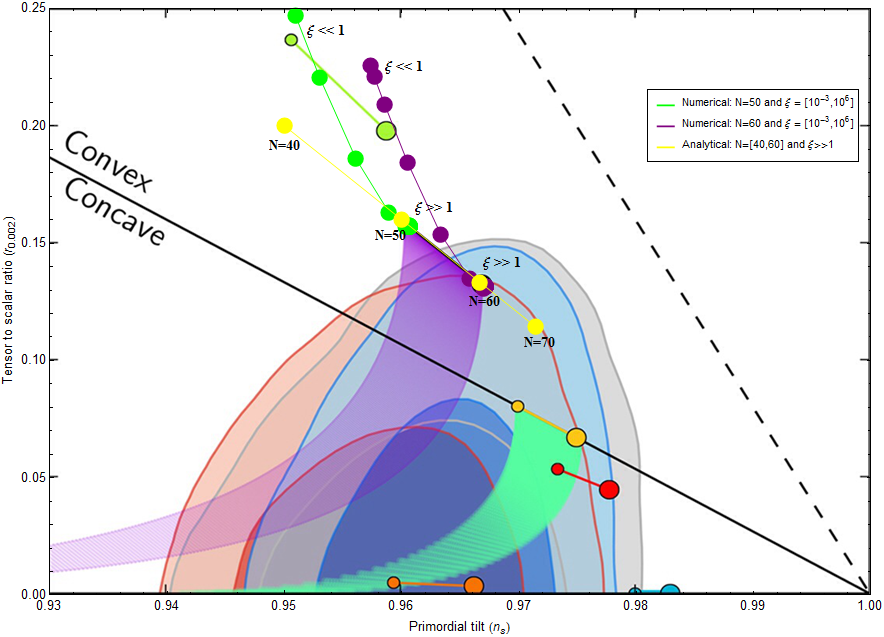}
\end{center}
\caption{The contours show the resulting 68$\%$ and 95$\%$ confidence regions for the tensor-to-scalar ratio $r$, the scalar spectral index $n_{s}$ and $N_{c}=1$. Left: The red contours are for the Planck+WP+highL data combination, while the blue ones display the BICEP2 constraints on $r$ \cite{Ade:2014xna}. Right: The figure shows the results from Planck plus various ancillary sets of data \cite{Ade:2013uln}. The plots also show the analytical and numerical predictions given by the sGB model.} 
\label{f1tosgb}
\end{figure}

The underlying gauge theory of this model is initiated in \cite{GVSY} based on the following considerations. Let us consider the pure $N=1$ supersymmetric Yang-Mills (SYM) gauge theory proposed by suitably modifying that of the ordinary QCD. The theory we are considering is the SU($N_{c}$) gauge group featuring a one flavor ($N_{f}=1$) gauge group with Weyl fermions in the adjoint representation. The Lagrangian can be written as 
\begin{eqnarray}
{\cal L}_{\rm SYM} = -\frac{1}{4}{\cal G}^{a}_{\mu\nu}{\cal G}^{a,\,\mu\nu}+\frac{i}{2}\bar{\lambda}^{a,\alpha}{\not}{D}_{ab}\lambda^{b}_{\alpha}+...\,,
\end{eqnarray}
where $\alpha$ is an ordinary spin, $a=1,\,.\,.\,.\,,N^{2}_{c}-1$, $\lambda^{a}$ is the spinor field and ${\cal G}^{a}_{\mu\nu},\,{\not}{D}_{ab}$ are the usual Yang-Mills strength tensor and a covariant derivative, respectively. The dots in principle represent \lq\lq gauge fixing, ghost terms and auxiliary fields\rq\rq~of those are not relevant for our current discussion. This theory is supersymmetric of an arbitrary $N_{c}$. If a strongly interacting regime takes place, the spinor fields (gluon fields) do condensate into a composite field, called super-glueball, which will be identified as the inflaton $\Phi$ in this case. The precise form of the inflaton field is prior given in \cite{Channuie:2012bv} such that $\Phi=-3\lambda^{a,\alpha}\lambda^{a}_{\alpha}/64\pi^{2}N_{c}$. As examined in the previous two examples, the fundamental degrees of freedom are naturally non-minimally coupled to gravity, and features the description at the fundamental level. We start with the introduction of the non-minimal coupling term as follows:
\begin{eqnarray}
\frac{N^{2}_{c}}{2}\xi\left(\frac{-3\lambda^{a,\alpha}\lambda^{a}_{\alpha}}{64\pi^{2}N_{c}}\right)^{2/3}R\equiv \frac{N^{2}_{c}\xi\Phi^{2/3}R}{2}\,.
 \end{eqnarray}
Again, the $\xi$ is the dimensionless coupling. We have just demonstrated how the introduction of the non-minimal coupling is motivated in a natural way with the underlying fundamental descriptions. According to this model, the inflaton is designed to be the gluino-ball state in the super--Yang--Mills theory. For this model, we have
\begin{eqnarray}
 f(\varphi)=4\alpha N^{2}_{c}\ln^{2}(\varphi/\Lambda)\,.
 \end{eqnarray}
where the gluino condensate is parametrised by $\Lambda$ and we find in \cite{Channuie:2012bv} that $\Lambda^{3}=(9/32\pi^{2})\Lambda^{3}_{\rm SUSY,\,YM}$.
 \begin{figure}
\begin{center}
\includegraphics[width=.6\textwidth]{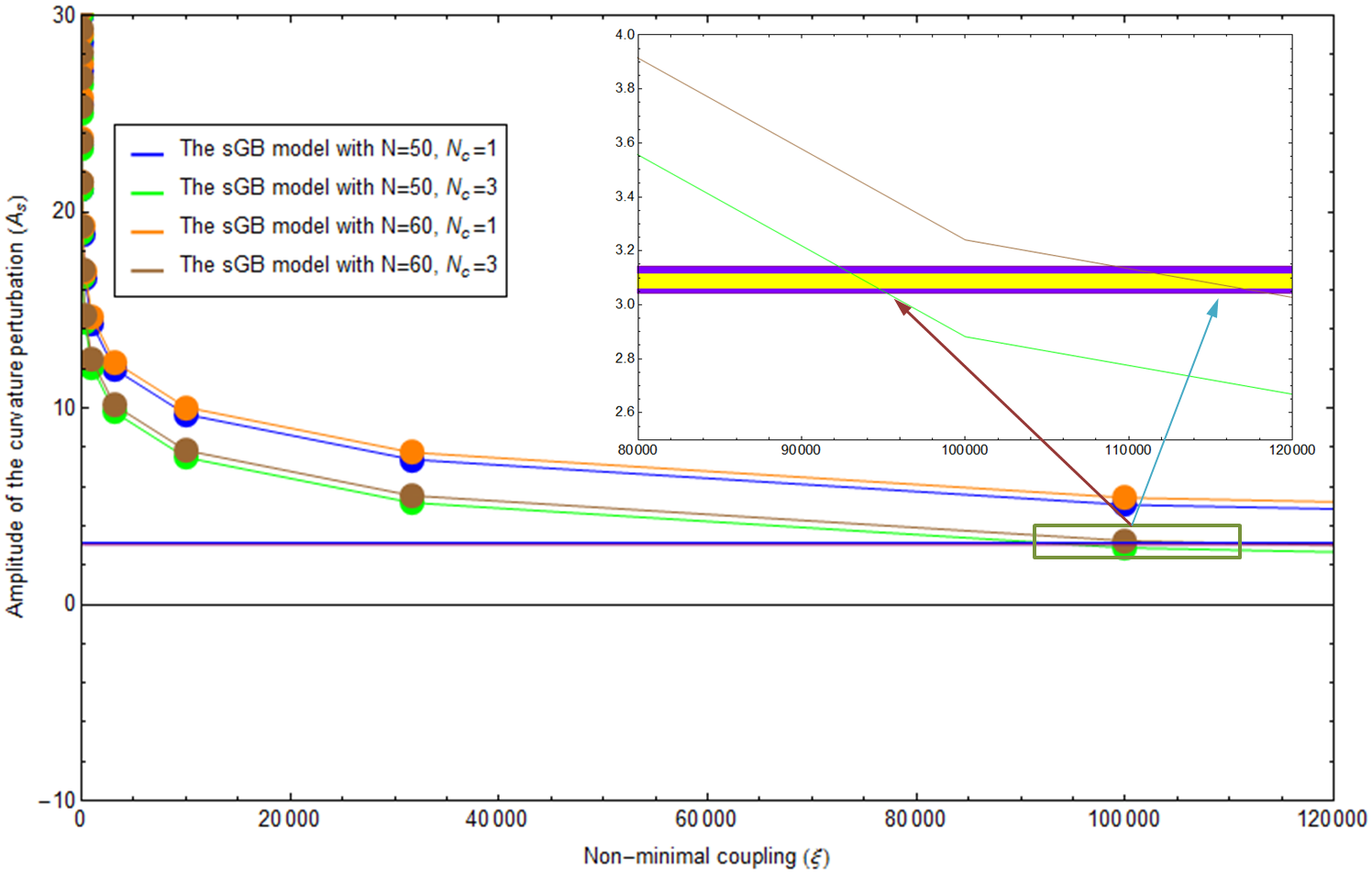} 
\end{center}
\caption{The plot shows the relation between the amplitude of the power spectrum ${\cal A}_{s}$ and the nonminimal coupling $\xi$ with $10^{-3}\lesssim \xi\lesssim 10^{6}$ for ${\cal N}=50,60$ predicted by the sGB model. The horizontal bands represent the $1\sigma$ (yellow) and $2\sigma$ (purple) C.L. for ${\cal A}_{s}$ obtained from Planck.} 
\label{f2sgb}
\end{figure}
As it is always investigated in standard fashion, we take the scalar component part of the superglueball action and coupled it nonminimally to gravity. Focusing only on the modulus of the inflaton field and taking the next step in order to write the non-minimally coupled scalar component part of the superglueball action to gravity, the resulting action in the Jordan frame reads \cite{Channuie:2013lla}
\begin{eqnarray}
{\cal S}_{\rm sGB} = \int d^{4}x\sqrt{-g}\Big[\frac{F(\varphi)}{2} R - \frac{9N^2_{c}}{2\alpha}g^{\mu\nu} \partial_{\mu}\varphi\partial_{\nu}\varphi - 4\alpha N^{2}_{c}\varphi^{4}(\ln[\varphi/\Lambda])^{2} \Big]\,,
\end{eqnarray}
with $N_{c}$ a number of colors, and $\alpha$ a $N_{c}$-independent quantity. Here we call it, in brief, the sGB model. Using the similar approximations to those of the above consideration, the number of e-foldings for this inflation model in the large $\xi$ limit is approximately given by
\ba
{\cal N} \simeq \frac{3}{2}\Big(\ln^2\(\varphi/\Lambda\) - \ln^2\(\varphi_{e}/\Lambda\)\Big) + {\cal O}(1/\xi)\,.
\label{efold-sgb-l}
\ea
Regarding to the above relations between the number of e-foldings and $\varphi$,
we can write $n_s$, $r$ and $|\zeta|^2$ in terms of ${\cal N}$ for a large $\xi$ limit to yield
\begin{eqnarray}
n_{s} 
&\simeq&  1 - \frac 2{{\cal N}} + {\cal O}(1/\xi)\,,
\label{ns-sgb-l}\\
r 
&\simeq&  \frac 8{{\cal N}} + {\cal O}(1/\xi)\,,
\label{t2s-sgb-l}\\
|\zeta|^2 &\simeq&  \frac{2\alpha{\cal N}^{2}}{81N^{2}_{c}\pi^{2}\xi^{2}} + {\cal O}(1/\xi^3)\,.
\label{zeta2-sgb-l}
\end{eqnarray}
The consistency relation of the above relations reads
\begin{eqnarray}
r \simeq 4(1-n_{s})\,.
\end{eqnarray}
We discover that the predictions of this model are fully consistent with BICEP2 constraints for ${\cal N} \subseteq [50, 60]$. Moreover, the model can also be consistent with the Planck contours at $1\sigma$ CL. We discover that ${\cal A}_{s}$ is well consistent with the Planck data up to $2\sigma$ CL for ${\cal N}=50$ and $N_{c}=3$ with $9.2\times 10^{4}\lesssim \xi \lesssim 9.5\times 10^{4}$, illustrated in Fig\,(\ref{f2sgb}). This model provides $n_{s}\simeq 0.960$ and $r\simeq 0.16$ for ${\cal N}=50$ with $\xi\gg 1$, see Fig.~(\ref{f1tosgb}). Here we can use the BICEP2 results to constrain $\Lambda_{\rm SgbI}$ since the data provides us the lower bound on $r$. According to the recent BICEP2 data, we roughly opt $r\simeq 0.12$ and use $N_{c}=1(3)$ predicting $\Lambda_{\rm sGB}> 10^{-3}(10^{-4})$ which corresponds to, at least, the GUT energy scale in this investigation, in order to satisfy the BICEP2 data at $1\sigma$ CL. We hope that the future observations will provide significant examination for this model. 

We will complete our discussion in this section by naively clarifying the scales of the theory. As we have seen for the MCI, the effective description of this model would also break down at some energy scale. It was found in \cite{Channuie:2012bv} that $\Lambda \sim (0.57/\sqrt{N_{c}})M_{\rm P}/\sqrt{\xi}$. This value is not only consistent with the results found in \cite{Channuie:2011rq,Bezrukov:2011mv} but also shows that it is possible to lower the scale of composite inflation by increasing the number of underlying colors, $N_{c}$. 

With the result given in \cite{Channuie:2012bv}, i.e. $\varphi_{\rm ini}\sim 570\Lambda$, we expect that the fundamental description can be used in the perturbative regime to describe the dynamics of the theory at energy scales of the order of $600\Lambda$ and above. For energies below this scale and to describe the vacuum properties of the theory, the effective potential utilised in this presentation works.

Since the standard model couplings are weak at the unification point, whilst the inflationary model is still strongly coupled at this scale (now identified with $\Lambda$). this feature allows us to decouple the contributions of the SM from inflationary theory and ensures that the action formulating inflation does not include any contributions from the SM. For this model, we show that inflation starts at energy scales just below or near the energy scales above which the underlying gauge dynamic is perturbative and expect the perturbative dynamic of the gauge theory to set in before arriving at the Planck scale.  

From Eq.~(\ref{Hub}), we can determine the Hubble parameter during inflation and roughly find that $H\lesssim M_{\rm P}/\xi\sim {\cal O}(10^{14})$\,GeV. Apparently, the Hubble scale during inflation is less than all scales we have in this model ensuring the validity of the effective theory during inflation.
\subsection{Composite Inflation from Orientifold Theory}

The authors of \cite{Channuie:2012bv} examined the supersymmetric low-energy effective action to study inflation driven by the gauge dynamics of SU(N) gauge theories adding one Dirac fermion in either the two-index antisymmetric or symmetric representation of the gauge group. Such theories are known as orientifold theories \cite{Sannino:2003xe}. Here the gluino field of supersymmetric gluodynamics is replaced by two Weyl fields which can be formed as one Dirac spinor. The background framework of this model is to slightly deform an effective Lagrangian for the pure $N=1$ supersymmetric Yang-Mills theory derived in \cite{Veneziano:1982ah}. For investigating the inflationary scenario, we write the action by using the real part of the field $\varphi$ in which the orientifold sector non-minimally coupled to gravity in the Jordan frame \cite{Channuie:2012bv}
\begin{eqnarray}
{\cal S}_{\rm OI,\,J}\supset \int d^{4}x\sqrt{-g}\left[-\frac{F(\varphi)}{2} R - \frac{9F(N_{c})}{\alpha}g^{\mu\nu} \partial_{\mu}\varphi\partial_{\nu}\varphi- 4\alpha F(N_{c})\varphi^{4}\left(\ln(\varphi/\Lambda)^{2} -\gamma\right)  \right]\,, 
\end{eqnarray}
where $F(\varphi)=1 + N^{2}_{c}\xi\,\varphi^{2}$, $F(N_{c})=N^{2}_{c}(1+{\cal O}(1/N_{c}))$, $\gamma=1/9N_{c}+{\cal O}(1/N^{2}_{c})$ and hereafter we will keep only leading order in $1/N_{c}$. However, we can impose the conformal transformation and then find the resulting action in the Einstein frame    
\begin{eqnarray}
{\cal S}_{\rm OI,\,E}\supset\int d^{4}x\sqrt{-g}\left[-\frac{M^{2}_{P}}{2} R - \frac{1}{2}g^{\mu\nu} \partial_{\mu}\chi\partial_{\nu}\chi- \frac{4\alpha F(N_{c})}{N^{4}_{c}}\frac{M^{4}_{P}}{\xi^{2}}\Big[\ln\Big(\varphi/\Lambda\Big)^{2}-\gamma\Big]  \right]\,, 
\end{eqnarray}
with $N_{c}$ being a number of colours. Note that at large-$N_{c}$ the theory features certain super Yang-Mills properties, i.e. $F(N_{c})\rightarrow N^{2}_{c}$. With this limit, the transformed potential reduces to that of Section~(\ref{sGB}). With the large field limit, we can derive the following slow-roll parameter in terms of the number of e-foldings as
\begin{eqnarray}
n_{s}=1-6\epsilon+2\eta\simeq 1-\frac{2}{{\cal N}}\left(1+\frac{9\gamma}{2{\cal N}}\right)\,,\,\,\,\,r=16\epsilon\simeq \frac{8}{{\cal N}}\left(1+\frac{3\gamma}{{\cal N}}\right)\,. \label{paraorient}
\end{eqnarray}
Notice that for large $N_{c}$ the observables given above features the Super Yang-Mills inflation since $\gamma\rightarrow 0$. We will complete our discussion in this section by naively clarifying the scales of the theory. Recall that the underlying theory of this model is just a deformation of the previous one. For this model, it was found in \cite{Channuie:2012bv} that the typical scale of $\Lambda $ is also the typical scale for grand unification. However, such a scale in this model contains additional modifications due to the presence of small parameters which are inversely promotional to the number of underlying colors.  However, such contributions are negligible compared with the scales themselves. Also the Hubble scale during inflation is less than all scales we have in this model ensuring the applicability of the effective theory during inflation.

\section{Conclusions}
\label{conclusions}

We revisit single-field (slow-roll) inflation in which the inflaton is a composite field stemming from various strongly interacting field theories. With regard to our framework, the cosmological \lq\lq hierarchy problem\rq\rq\, in the scalar sector of the inflation can effectively be solved, which not be solved by Higgs or another elementary scalar field paradigms. We constrain the number of e-foldings for composite models of inflation in order to obtain a successful inflation. We study a set of cosmological parameters, e.g., the primordial spectral index $n_{s}$ and tensor-to-scalar ratio $r$, and confront the predicted results with the joint Planck data, and with the recent BICEP2 data. Last but not the least, we anticipate that the composite paradigms and their verifiable consequences, e.g., reheating mechanism, can possibly receive considerable attention for inflationary model buildings.

\bigskip

\noindent {\bf Acknowledgments.}  P.C. is financially supported by the Thailand Research Fund (TRF) under the project of the \lq\lq TRF Grant for New Researcher\rq\rq\,with Grant No.\,TRG5780143.


\begin{thebibliography}{nn}


 \bibitem{Starobinsky:1979ty} 
  A.~A.~Starobinsky,
  JETP Lett.\  {\bf 30}, 682 (1979)
  [Pisma Zh.\ Eksp.\ Teor.\ Fiz.\  {\bf 30}, 719 (1979)].
  
  \bibitem{Starobinsky:1980te} 
  A.~A.~Starobinsky,
  Phys.\ Lett.\ B {\bf 91}, 99 (1980).
  
  \bibitem{Mukhanov:1981xt} 
  V.~F.~Mukhanov and G.~V.~Chibisov,
  JETP Lett.\  {\bf 33}, 532 (1981)
  [Pisma Zh.\ Eksp.\ Teor.\ Fiz.\  {\bf 33}, 549 (1981)].

  \bibitem{Guth:1980zm} 
  A.~H.~Guth,
  Phys.\ Rev.\ D {\bf 23}, 347 (1981).
  
  \bibitem{Linde:1981mu} 
  A.~D.~Linde,
  Phys.\ Lett.\ B {\bf 108}, 389 (1982).
  
  \bibitem{Albrecht:1982wi} 
  A.~Albrecht and P.~J.~Steinhardt,
  Phys.\ Rev.\ Lett.\  {\bf 48}, 1220 (1982).
  
  
\bibitem{Nakayama:2014koa} 
  K.~Nakayama and F.~Takahashi,
  Phys.\ Lett.\ B {\bf 734}, 96 (2014)

\bibitem{Cook:2014dga} 
  J.~L.~Cook, L.~M.~Krauss, A.~J.~Long and S.~Sabharwal,
  Phys.\ Rev.\ D {\bf 89}, 103525 (2014)
  
\bibitem{Hamada:2014iga} 
  Y.~Hamada, H.~Kawai, K.~y.~Oda and S.~C.~Park,
  Phys.\ Rev.\ Lett.\  {\bf 112}, 241301 (2014)
  
\bibitem{Germani:2014hqa} 
  C.~Germani, Y.~Watanabe and N.~Wintergerst,
  arXiv:1403.5766 [hep-ph].
  
\bibitem{Oda:2014rpa} 
  I.~Oda and T.~Tomoyose,
  arXiv:1404.1538 [hep-ph].
  
\bibitem{Harigaya:2014sua} 
  K.~Harigaya, M.~Ibe, K.~Schmitz and T.~T.~Yanagida,
  Phys.\ Lett.\ B {\bf 733}, 283 (2014)
  
\bibitem{Lee:2014spa} 
  H.~M.~Lee,
  arXiv:1403.5602 [hep-ph].
  
\bibitem{Harigaya:2014qza} 
  K.~Harigaya and T.~T.~Yanagida,
  arXiv:1403.4729 [hep-ph].
  
\bibitem{Czerny:2014qqa} 
  M.~Czerny, T.~Higaki and F.~Takahashi,
  Physics Letters B (2014) 167-172
  
\bibitem{Ellis:2014cma} 
  J.~Ellis, N.~E.~Mavromatos and D.~V.~Nanopoulos,
  Phys.\ Lett.\ B {\bf 732}, 380 (2014)
  
\bibitem{Viaggiu:2014moa} 
  S.~Viaggiu,
  arXiv:1403.2868 [astro-ph.CO].
  
\bibitem{Kehagias:2014wza} 
  A.~Kehagias and A.~Riotto,
  Phys.\ Rev.\ D {\bf 89}, 101301 (2014)
  
\bibitem{Kobayashi:2014jga} 
  T.~Kobayashi and O.~Seto,
  Phys.\ Rev.\ D {\bf 89}, 103524 (2014)
  
\bibitem{Hertzberg:2014aha} 
  M.~P.~Hertzberg,
  arXiv:1403.5253 [hep-th].
  
\bibitem{Okada:2014lxa} 
  N.~Okada, V.~N.~?eno?uz and Q.~Shafi,
  arXiv:1403.6403 [hep-ph].
  
\bibitem{Ferrara:2014ima} 
  S.~Ferrara, A.~Kehagias and A.~Riotto,
  Fortsch.\ Phys.\  {\bf 62}, 573 (2014)
  
\bibitem{Gong:2014cqa} 
  Y.~Gong and Y.~Gong,
  Phys.\ Lett.\ B {\bf 734}, 41 (2014)
  
\bibitem{Bamba:2014jia} 
  K.~Bamba, R.~Myrzakulov, S.~D.~Odintsov and L.~Sebastiani,
  arXiv:1403.6649 [hep-th].
  
\bibitem{DiBari:2014oja} 
  P.~Di Bari, S.~F.~King, C.~Luhn, A.~Merle and A.~Schmidt-May,
  arXiv:1404.0009 [hep-ph].
  
\bibitem{Palti:2014kza} 
  E.~Palti and T.~Weigand,
  JHEP {\bf 1404}, 155 (2014)
  
\bibitem{Kumar:2014oka} 
  K.~S.~Kumar, J.~Marto, N.~J.~Nunes and P.~V.~Moniz,
  JCAP {\bf 1406}, 064 (2014)
  
\bibitem{Fujita:2014iaa} 
  T.~Fujita, M.~Kawasaki and S.~Yokoyama,
  arXiv:1404.0951 [astro-ph.CO].
  
\bibitem{Chung:2014woa} 
  Y.~C.~Chung and C.~Lin,
  JCAP {\bf 1407}, 020 (2014)
  
\bibitem{Antusch:2014cpa} 
  S.~Antusch and D.~Nolde,
  JCAP {\bf 1405}, 035 (2014)
  
\bibitem{Bastero-Gil:2014oga} 
  M.~Bastero-Gil, A.~Berera, R.~O.~Ramos and J.~G.~Rosa,
  arXiv:1404.4976 [astro-ph.CO].
  
\bibitem{Kawai:2014doa} 
  S.~Kawai and N.~Okada,
  arXiv:1404.1450 [hep-ph].
  
\bibitem{Hossain:2014coa} 
  M.~. W.~Hossain, R.~Myrzakulov, M.~Sami and E.~N.~Saridakis,
  Phys.\ Rev.\ D {\bf 89}, 123513 (2014)
  [arXiv:1404.1445 [gr-qc]].
  
\bibitem{Kannike:2014mia} 
  K.~Kannike, A.~Racioppi and M.~Raidal,
  JHEP {\bf 1406}, 154 (2014)
  
\bibitem{Ho:2014xza} 
  C.~M.~Ho and S.~D.~H.~Hsu,
  JHEP {\bf 1407}, 060 (2014)
  
\bibitem{Joergensen:2014rya}
  J.~Joergensen, F.~Sannino and O.~Svendsen,
  Phys.\ Rev.\ D {\bf 90} (2014) 043509
  [arXiv:1403.3289 [hep-ph]].
  

\bibitem{Mortonson:2014bja}
  M.~J.~Mortonson and U.~Seljak,
  JCAP {\bf 1410} (2014) 10,  035
  [arXiv:1405.5857 [astro-ph.CO]].
  
\bibitem{Adam:2014bub}
  R.~Adam {\it et al.}  [Planck Collaboration],
  arXiv:1409.5738 [astro-ph.CO].
  
\bibitem{Mortonson:2014bja}
  M.~J.~Mortonson and U.~Seljak,
  JCAP {\bf 1410} (2014) 10,  035
  [arXiv:1405.5857 [astro-ph.CO]].
  
\bibitem{Cheng:2014pxa}
  C.~Cheng, Q.~G.~Huang and S.~Wang,
  arXiv:1409.7025 [astro-ph.CO].
  

\bibitem{Bezrukov:2008ej}
  F.~L.~Bezrukov, A.~Magnin and M.~Shaposhnikov,
  Phys.\ Lett.\ B {\bf 675} (2009) 88
  [arXiv:0812.4950 [hep-ph]].
  
\bibitem{Bezrukov:2008ut}
  F.~Bezrukov, D.~Gorbunov and M.~Shaposhnikov,
  JCAP {\bf 0906} (2009) 029
  [arXiv:0812.3622 [hep-ph]].
  
\bibitem{DeSimone:2008ei}
  A.~De Simone, M.~P.~Hertzberg and F.~Wilczek,
  Phys.\ Lett.\ B {\bf 678} (2009) 1
  [arXiv:0812.4946 [hep-ph]].
  
\bibitem{Barvinsky:2009fy}
  A.~O.~Barvinsky, A.~Y.~Kamenshchik, C.~Kiefer, A.~A.~Starobinsky and C.~Steinwachs,
  JCAP {\bf 0912} (2009) 003
  [arXiv:0904.1698 [hep-ph]].
  
\bibitem{GarciaBellido:2008ab}
  J.~Garcia-Bellido, D.~G.~Figueroa and J.~Rubio,
  Phys.\ Rev.\ D {\bf 79} (2009) 063531
  [arXiv:0812.4624 [hep-ph]].
  
\bibitem{Bezrukov:2009db}
  F.~Bezrukov and M.~Shaposhnikov,
  JHEP {\bf 0907} (2009) 089
  [arXiv:0904.1537 [hep-ph]].

\bibitem{Burgess:2009ea}
  C.~P.~Burgess, H.~M.~Lee and M.~Trott,
  JHEP {\bf 0909} (2009) 103
  [arXiv:0902.4465 [hep-ph]].
  
 
\bibitem{Sannino:2004qp} 
  F.~Sannino and K.~Tuominen,
  Phys.\ Rev.\ D {\bf 71}, 051901 (2005)
  [hep-ph/0405209].
  
\bibitem{Hong:2004td} 
  D.~K.~Hong, S.~D.~H.~Hsu and F.~Sannino,
  Phys.\ Lett.\ B {\bf 597}, 89 (2004)
  [hep-ph/0406200].
  
\bibitem{Dietrich:2005wk} 
  D.~D.~Dietrich, F.~Sannino and K.~Tuominen,
  Phys.\ Rev.\ D {\bf 73}, 037701 (2006)
  [hep-ph/0510217].
  
\bibitem{Dietrich:2005jn} 
  D.~D.~Dietrich, F.~Sannino and K.~Tuominen,
  Phys.\ Rev.\ D {\bf 72}, 055001 (2005)
  [hep-ph/0505059].
  
\bibitem{Channuie:2011rq} 
  P.~Channuie, J.~J.~Joergensen and F.~Sannino, 
  JCAP {\bf 1105}, 007 (2011)
  [arXiv:1102.2898 [hep-ph]].
  
\bibitem{Hill:1991jc} 
  C.~T.~Hill and D.~S.~Salopek, 
  Annals Phys.\  {\bf 213}, 21 (1992).
  

\bibitem{Bezrukov:2011mv}
  F.~Bezrukov, P.~Channuie, J.~J.~Joergensen and F.~Sannino,
  Phys.\ Rev.\ D {\bf 86} (2012) 063513
  [arXiv:1112.4054 [hep-ph]].


\bibitem{Kaiser:1994vs}
  D.~I.~Kaiser,
  Phys.\ Rev.\ D {\bf 52} (1995) 4295
  [astro-ph/9408044].
  
\bibitem{Faraoni:1999hp}
  V.~Faraoni and E.~Gunzig,
  Int.\ J.\ Theor.\ Phys.\  {\bf 38} (1999) 217
  [astro-ph/9910176].
  
\bibitem{Catena:2006bd}
  R.~Catena, M.~Pietroni and L.~Scarabello,
  Phys.\ Rev.\ D {\bf 76} (2007) 084039
  [astro-ph/0604492].
  
\bibitem{Artymowski:2013qua}
  M.~Artymowski, Y.~Ma and X.~Zhang,
  Phys.\ Rev.\ D {\bf 88} (2013) 10,  104010
  [arXiv:1309.3045 [gr-qc]].


\bibitem{GVSY} 
G.~Veneziano and S.~Yankielowicz, 
Phys.\ Lett.\ B {\bf 113}, 3 (1982)


\bibitem{Channuie:2012bv} 
  P.~Channuie, J.~J.~Jorgensen and F.~Sannino, 
  Phys.\ Rev.\ D {\bf 86}, 125035 (2012)
  [arXiv:1209.6362 [hep-ph]].
  
\bibitem{Sannino:2003xe}
  F.~Sannino and M.~Shifman,
  Phys.\ Rev.\ D {\bf 69} (2004) 125004
  [hep-th/0309252].
  
\bibitem{Veneziano:1982ah}
  G.~Veneziano and S.~Yankielowicz,
  Phys.\ Lett.\ B {\bf 113} (1982) 231.
  

\bibitem{Channuie:2013lla}
  K.~Karwan and P.~Channuie,
  JCAP {\bf 1406} (2014) 045
  [arXiv:1307.2880 [hep-ph]].
  
\bibitem{Channuie:2013xoa}
  P.~Channuie,
  Int.\ J.\ Mod.\ Phys.\ D {\bf 23} (2014) 1450070
  [arXiv:1312.7122 [gr-qc]].

\bibitem{Channuie:2014kda}
  P.~Channuie and K.~Karwan,
  Phys.\ Rev.\ D {\bf 90} (2014) 047303
  [arXiv:1404.5879 [astro-ph.CO]].
  
\bibitem{Svendse:2012dva}
  O.~Svendse, ``Natural Models of Inflation,''
  
\bibitem{Evans:2010tf}
  N.~Evans, J.~French and K.~y.~Kim,
  JHEP {\bf 1011} (2010) 145
  [arXiv:1009.5678 [hep-th]].

  

\bibitem{Ade:2013uln} 
  P.~A.~R.~Ade {\it et al.}  [Planck Collaboration],
  ``Planck 2013 results. XXII. Constraints on inflation,''
  arXiv:1303.5082 [astro-ph.CO].
  
\bibitem{Ade:2014xna} 
  P.~A.~R.~Ade {\it et al.}  [BICEP2 Collaboration], ``BICEP2 I: Detection Of B-mode Polarization at Degree Angular Scales,''
  arXiv:1403.3985 [astro-ph.CO].



\end{thebibliography}
\end{document}